\documentclass[letterpaper]{article} 
\usepackage{aaai2027} 
\usepackage{times}  
\usepackage{helvet}  
\usepackage{courier}  
\usepackage[hyphens]{url}  
\usepackage{graphicx} 
\usepackage{natbib}  
\usepackage{caption} 
\usepackage{amsmath, amssymb, amsthm}
\usepackage{bm}
\usepackage{booktabs}
\usepackage{multirow}
\usepackage{enumitem}
\usepackage{algorithm}
\usepackage{algorithmic}
\usepackage{colortbl}
\usepackage{tcolorbox}
\usepackage{xcolor} 
\tcbuselibrary{breakable, skins}

\newcommand{\toolex}{\textsc{ToolEX}}
\newcommand{\toolret}{\textsc{ToolRet}}
\newcommand{\toolde}{\textsc{Tool-DE}}
\newcommand{\toolexb}{\textsc{ToolEq}}
\newcommand{\skillret}{\textsc{SkillRet}}
\newcommand{\skillexb}{\textsc{SkillEq}}

\newcommand{\calC}{\mathcal{C}}
\newcommand{\calT}{\mathcal{T}}
\newcommand{\ndcg}{\mathrm{NDCG}}
\newcommand{\recall}{\mathrm{Recall}}
\newcommand{\comp}{\mathrm{Comp}}
\newcommand{\oldv}[1]{\textcolor{gray}{#1}}
\newcommand{\pair}[2]{\oldv{#1}/#2}
\newcommand{\deltav}[1]{\textbf{#1}}

\usepackage{newfloat}
\usepackage{listings}
\DeclareCaptionStyle{ruled}{labelfont=normalfont,labelsep=colon,strut=off} 
\floatstyle{ruled}
\newfloat{listing}{tb}{lst}{}
\floatname{listing}{Listing}

\newcounter{promptlisting}

\newtcolorbox{prompttextbox}[1]{
 enhanced,
 breakable,
 colback=white,
 colframe=black,
 boxrule=0.8pt,
 arc=1mm,
 left=2mm, right=2mm, top=0.5mm, bottom=0.5mm,
 before skip=2pt, after skip=2pt,
 title=\texttt{#1},
 colbacktitle=black, coltitle=white,
 fonttitle=\bfseries\footnotesize,
 boxed title style={colframe=black,colback=black,boxrule=0.8pt,arc=1mm},
 fontupper=\ttfamily\footnotesize
}

\title{Tool Retrievers Are Underestimated:\\ Annotation Expansion Reveals True Capability}
\author{
    Yanyu Zhu\textsuperscript{\rm 1},
    Chenheng Zhang\textsuperscript{\rm 2}\equalcontrib,
    Shaoshen Chen\textsuperscript{\rm 1}\equalcontrib,
    Hoilam Pao\textsuperscript{\rm 1},
    Yufei Zhang\textsuperscript{\rm 3},
    Jiajun Chai\textsuperscript{\rm 3},
    Dongnian Wang\textsuperscript{\rm 3},
    Zhaoyu Hu\textsuperscript{\rm 3},
    Guojun Yin\textsuperscript{\rm 3},
    Wei Lin\textsuperscript{\rm 3},
    and Hai-Tao Zheng\textsuperscript{\rm 1}\thanks{Corresponding author.}
}
\affiliations{
    Shenzhen International Graduate School, Tsinghua University \\
    Peking University \\
    Meituan, Beijing
}

\begin{document}
\maketitle






\begin{abstract}
In open-world scenarios with massive and evolving tool repositories, tool-augmented large language models rely on a retriever to surface relevant tools for a given query. Because such repositories often contain many tools that implement the same functionality, a single query can often be resolved by several distinct but functionally equivalent tool combinations, making the natural query-to-tool mapping inherently one-to-many. However, existing tool retrieval benchmarks annotate each query with a single relevant tool combination, collapsing this one-to-many mapping into a rigid one-to-one annotation and causing valid retrieved tools to be misjudged as failures.
To address this, we propose \textbf{\toolex{}} (Tool Equivalent eXpansion),
a framework that automatically discovers and annotates the tool
combinations functionally equivalent to the labeled ones.
Applied to the 7,360-query \toolde{} benchmark, \toolex{} finds that
67.9\% of sub-queries admit equivalent alternatives, expanding the
singular ground truth to an average of 5.3 valid combinations per query. Using the expanded benchmark \toolexb{}, we re-evaluate eight base retrievers and two fine-tuned variants; metrics on \toolexb{} rise substantially over \toolde{}, showing that one-to-one annotation systematically underestimates retrievers and that 30--47\% of the reported fine-tuning gain is an evaluation artifact rather than genuine improvement. Applying the same pipeline to skill retrieval on \skillret{} further confirms that the one-to-one problem extends beyond tool retrieval.

\end{abstract}


\begin{figure}[t]
  \centering
  \includegraphics[width=\columnwidth]{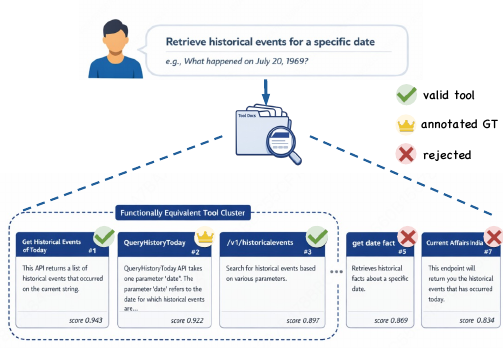}
  \caption{%
    \textbf{The one-to-many relationship between queries and tools in
    real-world tool libraries.}
    A single user query can be satisfied by multiple functionally equivalent
    tools, yet existing benchmarks annotate only one as ground truth.
  }
  \label{fig:one_to_many}
\end{figure}

\section{Introduction}
\label{sec:intro}

As LLM agents move toward open-domain deployment, tool retrieval becomes a
prerequisite for execution: an agent cannot invoke a capability it fails to
retrieve. This has motivated dedicated tool retrieval benchmarks such as
ToolRET~\citep{shi2025retrieval}, \toolde{}~\citep{lu2025tools}, and
MetaTool~\citep{huang2024metatoolbenchmarklargelanguage}. These benchmarks
are typically built by aggregating tools and task demonstrations from large
tool-use corpora, including ToolBench~\citep{qin2024toolllm},
ToolACE~\citep{liu2024toolace}, and ToolEyes~\citep{ye2024tooleyes}.
However, this construction process inevitably introduces tools with
overlapping functionality under different names, providers, or interfaces.
The resulting query-to-tool relation is therefore naturally one-to-many: a
single query may be solvable by multiple distinct but functionally equivalent
tool combinations. Existing benchmarks ignore this equivalence and instead
assign each query a single annotated ground-truth combination.
Consequently, a retriever that returns an equivalent but unannotated tool is
scored as wrong despite producing a valid retrieval. We call this the
\textbf{one-to-one annotation problem}.



Figure~\ref{fig:one_to_many} illustrates the problem on a concrete query that
asks to ``retrieve a list of historical events that occurred on a specific
calendar date.''
A dense retriever ranks candidate tools by similarity.
Only one tool, \texttt{QueryHistoryToday}, is annotated as the ground
truth; four others --- \texttt{Get Historical Events of Today},
\texttt{/v1/historicalevents}, \texttt{Historical Events by API Ninjas}, and
\texttt{v1\_historicalevents} --- are functionally equivalent yet unannotated,
and are therefore scored as misses, inducing a systematic underestimate of retrieval quality.



\begin{figure*}[t]
 \centering
 \includegraphics[width=\textwidth]{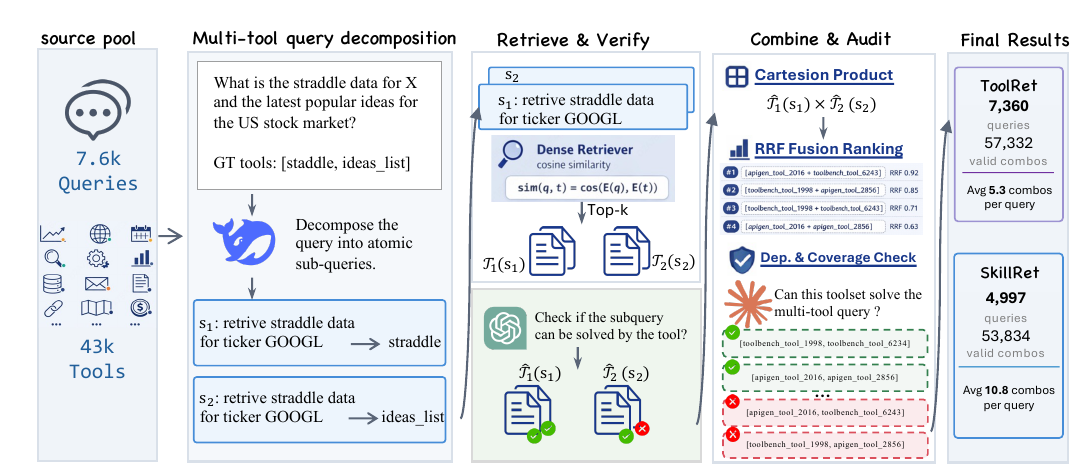}
 \caption{%
 \textbf{The \toolex{} annotation pipeline.}
 Stage~1 decomposes each query into atomic sub-queries via DeepSeek-V3.2;
 Stage~2 retrieves top-$K$ candidates per sub-query with Qwen3-Embedding-4B
 and verifies functional alignment via GPT-4o-mini;
 Stage~3 assembles Cartesian combinations, ranks them with RRF, and filters
 via Claude-Sonnet-5 dependency checking, yielding valid equivalent combinations.
 }
 \label{fig:pipeline}
\end{figure*}

To close this gap, we propose \toolex{}, a fully automated annotation pipeline
that discovers functionally equivalent tool combinations without human
labeling.
\toolex{} operates in three stages (Figure~\ref{fig:pipeline}):
an LLM first decomposes each composite query into $k$ atomic sub-queries, one
per tool in the ground-truth combination;
for each sub-query, a dense retriever (Qwen3-Embedding-4B~\citep{zhang2025qwen}) retrieves the top-20 candidate tools, which an LLM then verifies to determine whether they satisfy the capability specified by the sub-query; finally, the verified per-sub-query tool sets are merged via Cartesian product, ranked with Reciprocal Rank Fusion (RRF)~\citep{cormack2009reciprocal}, and filtered by an LLM dependency check that removes combinations violating inter-tool constraints. Applied to the 7,360-query \toolde{} evaluation set, which uses the same
queries as \toolret{}~\citep{shi2025retrieval} but augments the tool
documentation as in \toolde{}~\citep{lu2025tools}, \toolex{} expands
the average annotation from 1 to 5.3 tool combinations per query, with
67.9\% of sub-queries receiving at least one additional equivalent tool beyond
the original ground truth.

Using these expanded annotations, we construct \textbf{\toolexb{}}, a new
benchmark that credits a retrieval whenever it matches \emph{any} valid
equivalent combination, and re-evaluate eight base IR models and two fine-tuned tool retrievers.
\toolexb{} corrects a systematic underestimation: across the eight base
retrievers it raises NDCG@10 by 5--7\,pp on average over \toolde{}
(Table~\ref{tab:main}), recovering the credit withheld from retrievals of
functionally equivalent but unannotated tools.
This underestimate also inflates the apparent fine-tuning gain: correcting it narrows the gap between fine-tuned and base retrievers on \toolexb{} (0.6B: $+$6.4\,pp $\to$ $+$4.5\,pp; 4B: $+$7.2\,pp $\to$ $+$3.8\,pp), so part of the gain reported on \toolde{} reflects the evaluation protocol rather than only improved retrieval.
We further find that applying the same pipeline to the \skillret{} benchmark~\citep{cho2026skillret} confirms
the one-to-one annotation problem is ecosystem-wide.

In summary, our contributions are fourfold:
\begin{itemize}
  \item We unveil the one-to-one annotation problem in existing tool retrieval benchmarks, demonstrating that scoring functionally equivalent tools as misses systematically underestimates retriever capability.
  \item We propose \toolex{}, a fully automated three-stage pipeline that discovers functionally equivalent tool combinations without human annotation, and construct \textbf{\toolexb{}} as a new evaluation framework that credits any valid equivalent.
  \item By expanding singular ground truths to an average of 5.3 valid combinations per query, \toolexb{} corrects the underestimation: eight base retrievers gain 5--7\,pp NDCG@10 on average over \toolde{}. Correcting the metric also exposes that fine-tuning gains are partly an evaluation artifact, as the fine-tuned-to-base gap shrinks once equivalents are credited.
  \item Extending \toolex{} to \skillret{} shows the one-to-one problem is ecosystem-wide, confirming that functional equivalence is not specific to tool repositories.
\end{itemize}

\section{Related Work}
\label{sec:related}

\paragraph{Tool Learning Ecosystem: Use and Retrieval Benchmarks.}
Tool learning equips large language models (LLMs) with external tools, enabling them to execute actions and solve practical tasks as agents. 
A broad line of benchmarks evaluates tool-use capabilities across diverse execution scenarios, including ToolBench~\citep{qin2024toolllm}, APIBank~\citep{li2023apibank}, ToolAlpaca~\citep{qiao2024toolalpaca}, AppBench~\citep{yang2024appbench}, GTA~\citep{mao2024gta}, MetaTool~\citep{chang2024metatool}, ToolEyes~\citep{ye2024tooleyes}, ToolACE~\citep{liu2024toolace}, and APIGen~\citep{liu2024apigen}. 
However, in open-domain scenarios where tool/skill repositories scale to tens of thousands of items, feeding all documentation directly into the LLM context window becomes fundamentally infeasible. 
Tool and skill retrieval have thus emerged as indispensable prerequisites for open-domain agent execution, giving rise to dedicated benchmarks such as \toolret{}~\citep{shi2025retrieval}, \toolde{}~\citep{lu2025tools}, and \skillret{}~\citep{cho2026skillret}. 
To construct these retrieval benchmarks, standard pipelines collect large pools of candidate tools or skills, randomly sample seed tool subsets, and prompt LLMs to synthesize corresponding user queries. 
Crucially, this bottom-up synthesis pipeline inherently couples each generated query \emph{exclusively} with its seeded tools, creating an artificial \textbf{one-to-one annotation problem} that completely ignores functionally equivalent alternatives co-existing in the candidate repository.

\paragraph{Tool Retrieval Methodologies.}
To tackle the open-domain retrieval bottleneck, recent methodologies focus on optimizing retrieval quality along several complementary technical routes. 
A major line of work employs dense retriever fine-tuning via contrastive learning with negative sampling~\citep{karpukhin2020dense, xiong2021approximate}, as well as specialized vector space alignments~\citep{moon2024efficientscalableestimationtool} and generative identifiers~\citep{wang2025toolgenunifiedtoolretrieval}, to directly bridge the encoding gap between query intent and tool documentation. 
Complementary to representation learning, query reformulation and decomposition techniques~\citep{fang2026singleshotmultisteptoolretrieval, liu2025toolplannertaskplanningclusters, sengupta-etal-2026-tooldreamer, huang-etal-2024-planning} rewrite complex user intents into atomic sub-goals or leverage structural graph modeling~\citep{gao2025toolgraphretrieverexploring} to better match tool capabilities. 
More recently, joint optimization paradigms utilize reinforcement learning to co-optimize retrieval policies with agent execution in open-world environments~\citep{huang2026toolomni}. 
Crucially, while these approaches significantly advance \emph{how} queries and tools are encoded, reformulated, or retrieved, they universally evaluate predictions against a single, rigidly annotated target. 
As a result, whenever a model successfully surfaces a functionally equivalent alternative, traditional benchmarks falsely count it as a false negative. 
Our work is strictly orthogonal: rather than introducing another retrieval algorithm, we reform the evaluation paradigm itself by expanding singular ground truths into valid sets of functionally equivalent tool combinations.


\section{Equivalent Ground-truth for Tool Retrieval and Skill Retrieval}
\label{sec:method}

In this section, we develop \toolex{}, a framework that corrects the
one-to-one annotation problem in existing tool retrieval benchmarks. First, Section~\ref{subsec:limitations} analyzes the limitations of current benchmarks and formalizes the one-to-one annotation problem. Next, Section~\ref{subsec:pipeline} introduces the \toolex{} annotation pipeline for equivalent ground-truth expansion. Finally, Section~\ref{subsec:metric} presents the benchmarks \toolexb{} along with a new evaluation metric.

\subsection{Limitations of Current Retrieval Benchmarks}
\label{subsec:limitations}

Let $\mathcal{Q}$ denote a set of user queries and $\calT$ a heterogeneous
tool library.
Existing benchmarks~\citep{shi2025retrieval, lu2025tools} annotate each
query $q \in \mathcal{Q}$ with a single ground-truth tool combination
$C^* = \{t_1, \ldots, t_k\} \subseteq \calT$.
This one-to-one convention rests on a hidden assumption: that each query
admits exactly one correct tool combination.
In open-world repositories, this assumption fails.
Libraries assembled from independent sources inevitably contain
\emph{functionally equivalent} tools---APIs and services that perform the
same operation under different names, providers, or documentation
styles---so the true set of valid combinations for a query is
\begin{equation}
 \calC_q = \{ C \subseteq \calT \mid C \text{ can fully solve } q \},
\end{equation}
with $|\calC_q| \gg 1$ whenever the library contains equivalents.
Standard metrics treat only $C^*$ as relevant and count every
$C \in \calC_q \setminus \{C^*\}$ as a false negative, so the
query-to-relevant-tools relation is in reality one-to-many but is evaluated
as if it were one-to-one.
We call this the \textbf{one-to-one annotation problem}.

This mislabeling has two compounding consequences.
During retriever fitting, one-to-one annotations cause unlabeled positives
to contaminate the negative batch, which artificially tightens decision
boundaries and suppresses generalization to functionally equivalent tools.
On the \emph{evaluation} side, a retriever that correctly identifies an
unlabeled equivalent receives no credit, while one that overfits to the
single annotated combination is rewarded.
The two effects reinforce each other: measured gains on benchmarks such as
\toolde{} may therefore substantially overstate real improvement, and
there is currently no way to tell how much of a measured gain is genuine
retrieval ability and how much is benchmark artifact.
\toolex{} targets this structural source of bias by approximating
$\calC_q$ automatically, without human annotation.

\subsection{Annotation Pipeline}
\label{subsec:pipeline}

\toolex{} operates in three stages, as illustrated in
Figure~\ref{fig:pipeline}.
In the experiments reported here, the pipeline is applied to the
\toolde{} evaluation set derived from ToolRet~\citep{shi2025retrieval}
and \toolde{}~\citep{lu2025tools}, where ToolRet provides the
query--tool structure and \toolde{} provides expanded tool documentation
used for retrieval.

\paragraph{Stage 1: Multi-Tool Query Decomposition.}
The goal of this stage is to rewrite each composite query $q$ as $k$
atomic, provider-agnostic sub-queries $\{s_1, \ldots, s_k\}$, one per tool
in the ground-truth combination $C^*$, so that each $s_i$ can retrieve
functionally equivalent tools without revealing their names.
DeepSeek-V3.2~\citep{deepseekv3.2} performs the decomposition
(see supplementary materials for the full prompt), given the query, the task instruction, and the
documentation of all $k$ ground-truth tools, and is instructed to minimize
the semantic gap between each sub-query and its corresponding tool
documentation.
Because Stage~1 is used to construct expanded annotations, it intentionally
uses the benchmark-provided ground-truth tool set; it is therefore an
annotation-time component rather than a deployable test-time planner.
Each sub-query describes two aspects: (a)~the \emph{functionality} (the
precise operation, e.g., retrieve, calculate, search) and (b)~the
\emph{entity} (the parameter values the query context can supply, e.g., a
stock ticker, a calendar date).
The functionality term aligns the sub-query with equivalent tool
documentation, while the entity term lets the Stage~2 verifier confirm the
caller can supply the inputs the candidate tool requires.


\paragraph{Stage 2: Candidate Retrieval and Functional Verification.}
The goal of this stage is, for each sub-query $s_i$, to collect every tool
in $\calT$ that is functionally equivalent to the annotated ground-truth
tool $t_i$ and could substitute for it in solving $s_i$.
We first retrieve the top-$K$ candidate tools for $s_i$ with
Qwen3-Embedding-4B~\citep{zhang2025qwen}, then verify each candidate $t$
with GPT-4o-mini (full verifier prompt in the supplementary materials).
The verifier receives the sub-query $s_i$, the candidate's full
documentation as raw JSON (to avoid information loss from field parsing),
and the ground-truth tool's documentation as a reference, and judges
whether $t$ performs the same core operation as the reference and whether
its output covers what $s_i$ requires with parameters that $s_i$ can
supply.
A candidate is accepted only when both conditions hold; under genuine
uncertainty the verifier defaults to ``no'' to favor precision.
The verified set is
$\hat{\calT}_i = \{t \in \calT : \text{LLM verifies } t \text{ for } s_i\}$,
and by construction $t_i$ always passes verification (it is compared
against itself), guaranteeing $C^* \in$ the final expansion.

\paragraph{Stage 3: Combination Assembly and Audit.}
The goal of this stage is to assemble the per-sub-query verified sets
$\hat{\calT}_i$ into query-level tool combinations that genuinely solve
$q$, and to discard those that do not.
Candidate combinations are formed by Cartesian product,
\begin{equation}
 \hat{\calC}_q = \hat{\calT}_1 \times \hat{\calT}_2 \times \cdots
 \times \hat{\calT}_k,
\end{equation}
and ranked by Reciprocal Rank Fusion
(RRF~\citep{cormack2009reciprocal}) over per-sub-query retrieval scores.
Claude-Sonnet-5 then audits each combination (full audit prompt in
the supplementary materials) against the original query, the task instruction, and the sub-query
breakdown, with the ground-truth set as a reference, checking two
conditions:
\textbf{(1) Function coverage} --- the combination collectively performs
every operation the query requires, using the sub-query breakdown as a
checklist;
\textbf{(2) Dependency consistency} --- when one sub-query's output
(e.g., an authentication token, resource ID, or session handle) is
consumed by another, the corresponding tools come from the same platform,
since runtime values cannot cross platforms.
Combinations passing both checks form the valid equivalent set
$\hat{\calC}_q \supseteq \{C^*\}$.


\paragraph{Annotation Results.}
Table~\ref{tab:stats} summarizes the expansion annotation statistics on the
evaluation sets of \toolexb{} and \skillexb{}.

\begin{table}[t]
\centering
\caption{Annotation statistics produced by the \toolex{} pipeline.
Side-by-side statistics are reported for the evaluation sets of
\toolexb{} and \skillexb{}.}
\label{tab:stats}
\setlength{\tabcolsep}{4pt}
\footnotesize
\begin{tabular}{lcc}
\toprule
\textbf{Statistic} & \textbf{ToolEq} & \textbf{SkillEq} \\
\midrule
Queries annotated            & 7,360  & 4,997   \\
Sub-queries (Stage~1)        & 13,506 & 8,347   \\
Verified items (Stage~2)     & 57,332 & 43,934  \\
Avg.\ items / sub-query      & 4.2    & 5.26    \\
Valid combinations (Stage~3) & 39,087 & 53,834  \\
Avg.\ combos / query         & 5.3    & 10.8    \\
\bottomrule
\end{tabular}
\end{table}

On the \toolexb{} evaluation set, the pipeline verifies 4.2 equivalent
tools per sub-query on average, yielding 5.3 valid equivalent combinations
per query.
On \skillexb{}, the same pipeline verifies 5.26 equivalent skills per
sub-query and yields 10.8 valid combinations per query.
These results confirm that functional equivalence is widespread:
67.9\% of sub-queries admit at least one equivalent tool beyond the
annotated one, and 75.2\% of queries admit at least one equivalent
combination beyond the annotated ground truth.



\subsection{Benchmark and Evaluation Metric}
\label{subsec:metric}

Using the expanded annotations on the evaluation set, we construct
\textbf{\toolexb{}}, a new tool retrieval benchmark in which each query is
paired with a \emph{set of pre-annotated valid tool combinations} rather than
just one labeled answer.

Let $L_q$ be the ordered retrieval list for query $q$. For a single
reference combination $C$, write
$\mathrm{NDCG}(L_q,C)$, $\mathrm{Recall}(L_q,C)$, $\mathrm{Comp}(L_q,C)$
for the three standard IR metrics (defined in \S\ref{subsec:setup}).
Given the expanded reference set $\hat{\calC}_q=\{C_1,\dots,C_M\}$ of valid
equivalent combinations, \toolexb{} treats every $C_j \in \hat{\calC}_q$ as a
labeled ground-truth answer for $q$. Evaluation then applies the original
one-to-one metric to each labeled reference separately and keeps the
best score for \emph{that same metric}:
\begin{equation}
\begin{aligned}
 \ndcg^{\,\toolexb}_q   &= \max_{C\,\in\,\hat{\calC}_q}\; \ndcg(L_q,\,C),\\
 \recall^{\,\toolexb}_q &= \max_{C\,\in\,\hat{\calC}_q}\; \recall(L_q,\,C),\\
 \comp^{\,\toolexb}_q   &= \max_{C\,\in\,\hat{\calC}_q}\; \comp(L_q,\,C).
\end{aligned}
\label{eq:metric}
\end{equation}
Operationally, this is a \emph{best-match over pre-annotated ground truths},
not a joint maximization across metrics. For each query, we (i) regard all
$C_j \in \hat{\calC}_q$ as valid labeled references, (ii) compute the
standard one-to-one metric against each $C_j$, and (iii) for NDCG, Recall,
and Comp \emph{separately}, report the highest score obtained against any
labeled equivalent combination. A retrieval is therefore credited as soon as
its ranked list best matches one of the pre-annotated valid alternatives,
instead of being penalized for missing the benchmark's originally chosen
combination.

\begin{table*}[t]
\centering
\caption{%
 Main results pairing the one-to-one benchmark \toolde{} against the
 new benchmark \textbf{\toolexb{}} (one-to-many, max-aggregation).
 All values are in \%; each metric cell reports \emph{old / new}, i.e.\
 \textcolor{gray}{\toolde{} (gray)} / \toolexb{} (black).
 $\Delta$ is the mean gain (\toolexb{} $-$ \toolde{}) averaged over
 N@10/R@10/C@10 within each category (pp);
 $^\dagger$~oracle-assisted sub-query diagnostic with RRF.
}
\label{tab:main}
\footnotesize
\setlength{\tabcolsep}{2.5pt}
\resizebox{\textwidth}{!}{%
\begin{tabular}{lcccc cccc cccc}
\toprule
\textbf{Model} &
\multicolumn{4}{c}{\textbf{Code}} &
\multicolumn{4}{c}{\textbf{Web}} &
\multicolumn{4}{c}{\textbf{Customized}} \\
\cmidrule(lr){2-5}\cmidrule(lr){6-9}\cmidrule(lr){10-13}
& N@10 & R@10 & C@10 & \deltav{$\Delta$}
& N@10 & R@10 & C@10 & \deltav{$\Delta$}
& N@10 & R@10 & C@10 & \deltav{$\Delta$} \\
\midrule
\multicolumn{13}{l}{\textit{Query-level}} \\
\quad BM25s
 & \pair{45.2}{49.5} & \pair{58.6}{62.7} & \pair{57.2}{61.3} & \deltav{+4.2}
 & \pair{28.5}{38.1} & \pair{35.9}{45.8} & \pair{24.0}{30.1} & \deltav{+8.5}
 & \pair{44.4}{48.6} & \pair{50.1}{54.1} & \pair{39.0}{42.1} & \deltav{+3.8} \\
\quad gte-Qwen2-1.5B
 & \pair{41.4}{46.0} & \pair{53.2}{56.9} & \pair{51.2}{54.9} & \deltav{+4.0}
 & \pair{37.1}{43.5} & \pair{47.3}{53.0} & \pair{29.8}{33.8} & \deltav{+5.4}
 & \pair{48.1}{53.9} & \pair{56.6}{62.0} & \pair{43.3}{48.0} & \deltav{+5.3} \\
\quad e5-mistral-7b
 & \pair{41.9}{47.4} & \pair{55.9}{60.4} & \pair{53.9}{58.4} & \deltav{+4.8}
 & \pair{31.9}{39.7} & \pair{41.4}{48.9} & \pair{27.5}{32.6} & \deltav{+6.8}
 & \pair{43.6}{49.3} & \pair{50.8}{58.4} & \pair{39.3}{46.0} & \deltav{+6.7} \\
\quad GritLM-7B
 & \pair{24.0}{27.5} & \pair{33.4}{37.4} & \pair{32.0}{36.1} & \deltav{+3.9}
 & \pair{29.5}{36.2} & \pair{38.1}{45.6} & \pair{24.4}{29.7} & \deltav{+6.5}
 & \pair{41.8}{46.6} & \pair{49.1}{54.8} & \pair{37.6}{42.7} & \deltav{+5.2} \\
\quad NV-Embed-v1
 & \pair{48.8}{52.1} & \pair{63.0}{63.2} & \pair{60.0}{63.5} & \deltav{+3.4}
 & \pair{32.5}{38.1} & \pair{39.6}{45.6} & \pair{24.0}{28.8} & \deltav{+5.5}
 & \pair{39.6}{44.5} & \pair{45.6}{51.1} & \pair{34.4}{39.2} & \deltav{+5.1} \\
\quad Qwen3-Embedding-0.6B
 & \pair{48.4}{53.8} & \pair{62.3}{66.0} & \pair{59.9}{63.6} & \deltav{+4.3}
 & \pair{37.3}{45.2} & \pair{46.3}{52.9} & \pair{29.4}{33.8} & \deltav{+6.3}
 & \pair{42.5}{51.7} & \pair{48.9}{58.3} & \pair{39.1}{47.2} & \deltav{+8.9} \\
\quad Qwen3-Embedding-4B
 & \pair{53.5}{60.2} & \pair{70.7}{74.2} & \pair{69.2}{72.5} & \deltav{+4.5}
 & \pair{38.7}{46.7} & \pair{47.9}{54.9} & \pair{30.4}{35.5} & \deltav{+6.7}
 & \pair{42.4}{53.1} & \pair{50.9}{61.7} & \pair{40.2}{49.0} & \deltav{+10.1} \\
\quad Qwen3-Embedding-8B
 & \pair{52.6}{59.1} & \pair{68.1}{71.1} & \pair{66.2}{69.0} & \deltav{+4.1}
 & \pair{40.8}{49.1} & \pair{49.5}{56.7} & \pair{31.9}{37.1} & \deltav{+6.9}
 & \pair{43.9}{54.7} & \pair{52.5}{62.1} & \pair{41.9}{49.9} & \deltav{+9.5} \\
\quad Tool-Embed-0.6B
 & \pair{52.1}{56.0} & \pair{65.7}{67.5} & \pair{64.0}{65.8} & \deltav{+2.5}
 & \pair{42.3}{49.1} & \pair{52.5}{58.1} & \pair{35.6}{39.7} & \deltav{+5.5}
 & \pair{53.1}{59.0} & \pair{61.8}{65.9} & \pair{47.6}{50.4} & \deltav{+4.3} \\
\quad Tool-Embed-4B
 & \pair{55.6}{59.9} & \pair{70.6}{72.6} & \pair{68.7}{70.7} & \deltav{+2.8}
 & \pair{44.6}{51.3} & \pair{54.5}{60.2} & \pair{37.4}{41.8} & \deltav{+5.6}
 & \pair{55.9}{60.2} & \pair{62.1}{65.5} & \pair{47.5}{50.2} & \deltav{+3.5} \\
\midrule
\multicolumn{13}{l}{\textit{Sub-query-level ($^\dagger$)}} \\
\quad BM25s$^\dagger$
 & \pair{71.6}{80.2} & \pair{83.6}{88.7} & \pair{80.8}{85.7} & \deltav{+6.2}
 & \pair{32.0}{40.5} & \pair{41.8}{49.4} & \pair{29.6}{34.8} & \deltav{+7.1}
 & \pair{50.6}{55.7} & \pair{58.6}{62.9} & \pair{46.8}{50.3} & \deltav{+4.3} \\
\quad gte-Qwen2-1.5B$^\dagger$
 & \pair{74.1}{82.7} & \pair{87.2}{90.5} & \pair{85.0}{88.2} & \deltav{+5.0}
 & \pair{37.7}{45.3} & \pair{49.2}{55.9} & \pair{33.9}{39.1} & \deltav{+6.5}
 & \pair{56.8}{62.7} & \pair{64.8}{69.6} & \pair{50.8}{54.4} & \deltav{+4.8} \\
\quad e5-mistral-7b$^\dagger$
 & \pair{74.2}{83.1} & \pair{85.8}{90.1} & \pair{82.4}{86.8} & \deltav{+5.9}
 & \pair{30.7}{40.0} & \pair{41.2}{50.3} & \pair{29.7}{35.7} & \deltav{+8.1}
 & \pair{49.7}{55.9} & \pair{58.9}{64.6} & \pair{46.3}{50.5} & \deltav{+5.4} \\
\quad GritLM-7B$^\dagger$
 & \pair{78.3}{86.4} & \pair{90.5}{93.1} & \pair{87.5}{90.3} & \deltav{+4.5}
 & \pair{33.3}{42.5} & \pair{43.2}{52.1} & \pair{31.2}{37.5} & \deltav{+8.1}
 & \pair{57.3}{62.7} & \pair{66.2}{70.4} & \pair{53.3}{56.0} & \deltav{+4.1} \\
\quad NV-Embed-v1$^\dagger$
 & \pair{78.5}{87.8} & \pair{91.6}{94.9} & \pair{88.3}{91.5} & \deltav{+5.3}
 & \pair{31.7}{41.1} & \pair{41.0}{50.1} & \pair{29.0}{35.4} & \deltav{+8.3}
 & \pair{57.3}{64.0} & \pair{64.9}{69.3} & \pair{51.6}{54.6} & \deltav{+4.7} \\
\quad Qwen3-Embedding-0.6B$^\dagger$
 & \pair{76.2}{84.7} & \pair{88.5}{92.0} & \pair{84.5}{88.1} & \deltav{+5.2}
 & \pair{38.5}{47.1} & \pair{50.1}{57.3} & \pair{34.3}{39.6} & \deltav{+7.0}
 & \pair{48.1}{56.8} & \pair{56.6}{64.4} & \pair{48.2}{53.7} & \deltav{+7.3} \\
\quad Qwen3-Embedding-4B$^\dagger$
 & \pair{76.5}{86.5} & \pair{91.0}{94.2} & \pair{88.1}{91.3} & \deltav{+5.5}
 & \pair{38.6}{47.6} & \pair{48.8}{56.3} & \pair{33.9}{39.3} & \deltav{+7.3}
 & \pair{49.8}{59.7} & \pair{59.6}{67.9} & \pair{49.2}{55.9} & \deltav{+8.3} \\
\quad Qwen3-Embedding-8B$^\dagger$
 & \pair{75.4}{83.8} & \pair{87.8}{90.7} & \pair{81.3}{84.1} & \deltav{+4.7}
 & \pair{48.0}{61.7} & \pair{59.9}{70.4} & \pair{48.7}{58.8} & \deltav{+11.5}
 & \pair{46.9}{55.8} & \pair{55.1}{62.8} & \pair{39.5}{44.9} & \deltav{+7.3} \\
\quad Tool-Embed-0.6B$^\dagger$
 & \pair{81.2}{87.1} & \pair{90.8}{92.4} & \pair{86.6}{88.2} & \deltav{+3.0}
 & \pair{53.5}{65.4} & \pair{65.7}{73.9} & \pair{53.8}{61.5} & \deltav{+9.3}
 & \pair{49.4}{55.7} & \pair{56.1}{61.4} & \pair{38.3}{41.9} & \deltav{+5.1} \\
\quad Tool-Embed-4B$^\dagger$
 & \pair{82.1}{87.2} & \pair{90.8}{92.5} & \pair{86.4}{88.2} & \deltav{+2.8}
 & \pair{54.2}{65.8} & \pair{67.0}{75.3} & \pair{55.4}{63.4} & \deltav{+9.3}
 & \pair{54.4}{60.7} & \pair{61.4}{66.3} & \pair{42.3}{45.7} & \deltav{+4.9} \\
\bottomrule
\end{tabular}
}
\end{table*}

\section{Experiments}
\label{sec:experiments}

This section first presents the experimental settings in
\S\ref{subsec:setup}, followed by the main results in
\S\ref{subsec:results}, a downstream task evaluation on
ToolBench~\citep{qin2024toolllm} in \S\ref{subsec:downstream} to
verify that \toolex{}-expanded annotations improve end-to-end agent
performance, and an extended experiment on skill retrieval
in \S\ref{subsec:skill}.

\subsection{Experimental Setup}
\label{subsec:setup}

All models are evaluated on both \toolde{}~\citep{lu2025tools} (original,
one-to-one annotation) and \textbf{\toolexb{}} (multi-set,
max-aggregation metric from Equation~\ref{eq:metric}).

\paragraph{Retrieval Methods.}
We evaluate two retrieval methods that differ in how queries are formulated
and results are aggregated:
\begin{itemize}[leftmargin=*, itemsep=1pt]
 \item \textbf{Query-level:} the multi-tool query is encoded and matched
 against tool documents as a single retrieval unit.
 This is the standard approach used by prior benchmarks.
 \item \textbf{Sub-query-level diagnostic ($^\dagger$):} we reuse
 the Stage~1 decomposition from the annotation pipeline to study how
 atomic capability descriptions affect ranking. Because this decomposition
 uses the benchmark-provided tool count and reference tools, it is an
 oracle-assisted upper bound rather than a deployable retrieval setting.
 Each sub-query retrieves an independent top-$K$ candidate list;
 the per-sub-query results are then merged via Reciprocal Rank Fusion
 (RRF) to produce the final query-level ranking. We set $K=20$ for each sub-query, yielding a maximum of $k \times 20$ candidates.
\end{itemize}

\paragraph{Baselines.}
\label{subsubsec:baselines}
We evaluate \textsc{Tool-Embed}-\{0.6B, 4B\}, trained with \toolde{}
data and eight representative retrieval models on both \toolde{} and \toolexb{}:
\begin{itemize}[leftmargin=*, itemsep=1pt]
 \item \textbf{Sparse retriever:} BM25s~\citep{bm25s}.
 \item \textbf{Dense retrievers:} gte-Qwen2-1.5B-instruct~\citep{gteqwen2},
 e5-mistral-7b-instruct~\citep{e5mistral}, GritLM-7B~\citep{gritlm},
 NV-Embed-v1~\citep{nvembed}, and
 Qwen3-Embedding~\citep{zhang2025qwen} (0.6B, 4B, 8B).
\end{itemize}

\paragraph{Metrics.}
We adopt three widely used IR metrics to evaluate tool retrieval performance:
(i) \emph{NDCG@$K$} ($N@K$), which considers both the relevance of retrieved
tools and their ranking positions;
(ii) \emph{Recall@$K$} ($R@K$), which measures the proportion of target tools
successfully retrieved within the top-$K$ results;
and (iii) \emph{Comprehensiveness@$K$} ($C@K$), which assigns $C@K = 1$ if
all target tools are included in the top-$K$ results and $0$ otherwise.
We report $K = 10$ per query category (Code, Web, Customized) and average.
On \toolexb{}, scores follow the max-aggregation metric
(Equation~\ref{eq:metric}).

\subsection{Main Results}
\label{subsec:results}




\subsubsection{A Systematic Underestimation.}
Table~\ref{tab:main} pairs \toolde{} (gray) with \toolexb{} (black) for eight base retrievers under query-level retrieval. All eight score higher under \toolexb{}: expanding the single ground truth with its equivalent combinations restores credit for retrievals the one-to-one metric had counted as misses. Averaged over Code, Web, and Customized, the NDCG@10 gap between the two metrics runs from $+$4.6\,pp (NV-Embed-v1) to $+$8.5\,pp (Qwen3-Embedding-8B), mean $+$6.5\,pp; BM25s, e5-mistral-7b, gte-Qwen2-1.5B, and GritLM-7B each show a 5.0--6.3\,pp gap. Each retriever's top-$K$ list already contains functionally equivalent tools, but \toolde{} leaves them unlabeled and so never scores them; \toolexb{} corrects this by crediting any equivalent combination (Equation~\ref{eq:metric}). The old-new gap therefore measures the very underestimation \toolexb{} recovers.

\subsubsection{Category-Level Heterogeneity.}
False-negative correction differs markedly by query category.
Using the Qwen3-Embedding-0.6B \toolexb{} vs.\ \toolde{} NDCG@10 delta as a proxy
for annotation recovery, the Customized category benefits most ($+$9.2\,pp:
42.5 $\to$ 51.7), Web moderately ($+$7.9\,pp: 37.3 $\to$ 45.2),
and Code the least ($+$5.4\,pp: 48.4 $\to$ 53.8).
The gradient reflects domain-specific redundancy: customized and web tool
libraries aggregate numerous functionally similar APIs across providers,
while code repositories (e.g., model hubs) tend to expose more distinctive
interfaces.


\subsubsection{Unlocking Retrieval Gains with Query Decomposition}
We further report a controlled sub-query decomposition diagnostic
($^\dagger$) that reuses the Stage~1 sub-queries from the annotation
pipeline, while keeping the retriever unchanged and retrieving over the
full tool library.
Under this setting, zero-shot Qwen3-Embedding-0.6B$^\dagger$ gains
12.6\,pp NDCG@10 over its composite-query counterpart on \toolexb{}
(averaged across categories). Tool-Embed-4B$^\dagger$
further reaches 70.2 NDCG@10.
These results suggest that current retrievers already encode much of the
required tool semantics, and that a large share of the remaining error comes
from semantic interference within composite queries rather than from a lack
of retriever capacity. In other words, better query decomposition can unlock
large gains without changing retriever parameters. We view this as a controlled estimate of the headroom from better query decomposition.

\subsubsection{The Illusion of Fine-Tuning Improvements.}

\begin{figure}[t]
 \centering
 \includegraphics[width=\columnwidth]{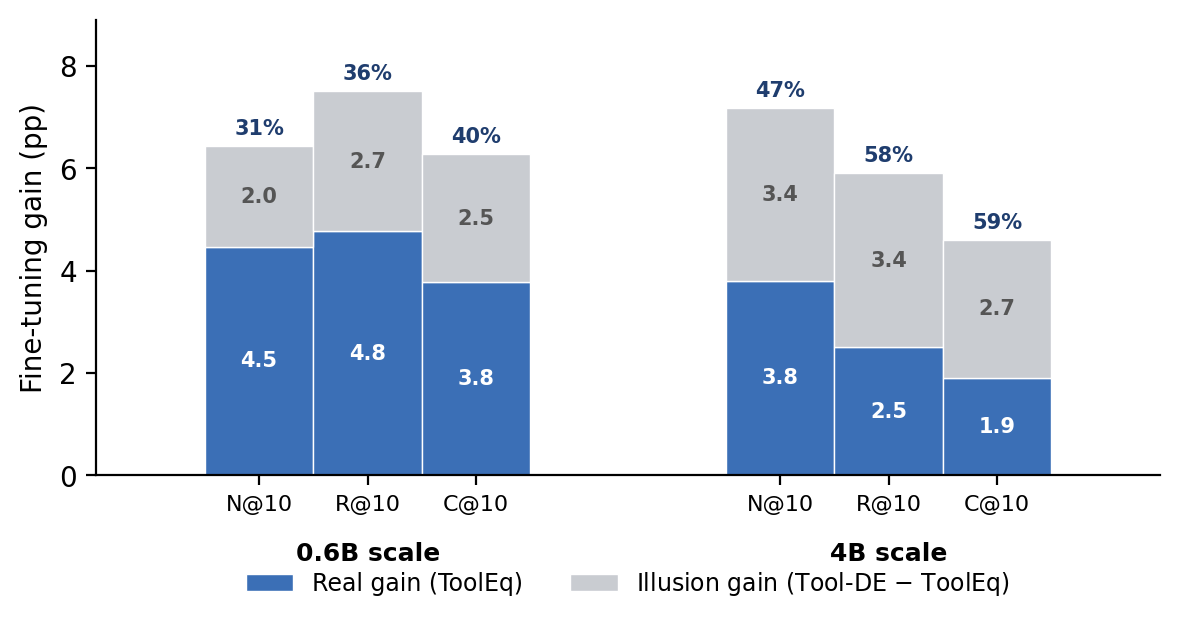}
 \caption{%
 Training gain illusion: The gray excess over blue is the illusion gain --- 47\% of the reported
 improvement at 4B is an evaluation artifact.
 }
 \label{fig:gain_illusion}
\end{figure}

We find that single-target evaluation can substantially overstate the
measured benefit of fine-tuning retrieval models. When evaluated on the
original \toolde{} benchmark, fine-tuning yields apparent gains:
Tool-Embed-0.6B improves over Qwen3-Embedding-0.6B by 6.4\,pp in NDCG@10
(averaged over Code/Web/Customized), and the 4B model by 7.2\,pp.
However, under \toolexb{} the same gaps shrink to 4.5\,pp at 0.6B and
3.8\,pp at 4B (Figure~\ref{fig:gain_illusion}). The inflation is most
pronounced at 4B, where 47\% of the reported gain vanishes once equivalents
are credited; on Code the sign even reverses (Tool-DE $+$2.1\,pp vs.\
ToolEq $-$0.3\,pp). This pattern indicates that a non-trivial fraction of
the measured gain comes from alignment to the single annotated target that
\toolde{} privileges. 



\subsubsection{Where Do Equivalent Tools Appear in the Ranking?}
Equivalent tools are often retrieved but left uncredited by one-to-one
evaluation. In the Qwen3-Embedding-4B ranking, the best-matching equivalent
tool appears in the top-10 for 79.4\% of queries with at least one verified
equivalent, slightly above the GT tool itself (76.1\%), while a random
negative reaches only 9.2\% on the same query set. Most importantly, for
14.5\% of these queries the GT tool falls outside the top-10 while an
equivalent tool appears within it. These are exactly the retrievals that
\toolde{} counts as failures but \toolexb{} correctly credits.

\begin{figure}[t]
 \centering
 \includegraphics[width=0.7\columnwidth]{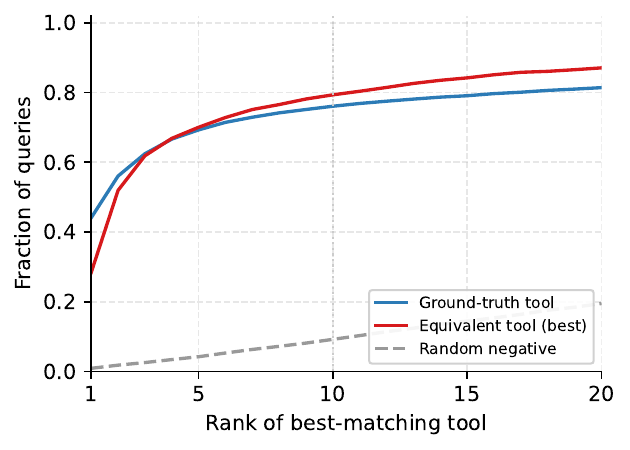}
 \caption{%
 CDF of the best rank of equivalent tools.}
 \label{fig:rank_cdf}
\end{figure}

\subsection{Extended Experiment: Skill Retrieval}
\label{subsec:skill}

To verify the generality of our framework beyond tools, we apply the
\toolex{} pipeline (Section~\ref{subsec:pipeline}) to skill retrieval on \skillret{}~\citep{cho2026skillret}. The merged expansion annotation statistics for
ToolEq and SkillEq are listed in Table~\ref{tab:stats}. We evaluate SkillRet-Embedding-{0.6B, 8B},
fine-tuned on \skillret{} training data, alongside eight retrieval models
detailed in Section~\ref{subsubsec:baselines}.
Table~\ref{tab:skill} reports the performance pairing
\skillret{} (one-to-one) against \textbf{\skillexb{}} (one-to-many).

\begin{table}[t]
\centering
\caption{%
  Skill retrieval on the full library (17,810 skills), pairing
  \skillret{} (one-to-one) against \textbf{\skillexb{}} (one-to-many).
  All values are in \%; each metric cell reports \emph{old / new}, i.e.,
  \textcolor{gray}{\skillret{} (gray)} / \skillexb{} (black).
  \deltav{$\Delta$} is the mean gain \skillexb{}$- $\skillret{} averaged over
  N@10/R@10/C@10 (pp).
}
\label{tab:skill}
\footnotesize
\setlength{\tabcolsep}{3pt}
\resizebox{\columnwidth}{!}{%
\begin{tabular}{lcccc}
\toprule
\textbf{Model} & N@10 & R@10 & C@10 & \deltav{$\Delta$} \\
\midrule
BM25s
  & \pair{42.0}{53.9} & \pair{49.4}{62.6} & \pair{34.3}{46.0} & \deltav{+12.3} \\
e5-mistral-7b
  & \pair{43.0}{53.8} & \pair{50.8}{61.8} & \pair{35.3}{44.9} & \deltav{+10.5} \\
gte-Qwen2-1.5B
  & \pair{49.1}{62.4} & \pair{57.1}{69.4} & \pair{41.0}{52.1} & \deltav{+12.2} \\
GritLM-7B 
  & \pair{54.1}{66.7} & \pair{62.3}{73.4} & \pair{46.1}{57.1} & \deltav{+11.6} \\
NV-Embed-v1 
  & \pair{57.6}{66.1} & \pair{65.2}{72.6} & \pair{48.7}{55.9} & \deltav{+7.7} \\
Qwen3-Embedding-0.6B
  & \pair{49.9}{60.8} & \pair{56.7}{66.3} & \pair{39.7}{48.6} & \deltav{+9.8} \\
Qwen3-Embedding-8B
  & \pair{50.2}{61.9} & \pair{57.1}{67.2} & \pair{39.9}{49.2} & \deltav{+10.4} \\
SkillRet-0.6B
  & \pair{53.2}{64.6} & \pair{61.1}{72.6} & \pair{45.4}{56.9} & \deltav{+11.5} \\
SkillRet-8B
  & \pair{74.6}{82.9} & \pair{83.2}{89.5} & \pair{72.7}{81.1} & \deltav{+7.7} \\
\bottomrule
\end{tabular}
}
\end{table}

\paragraph{One-to-Many Generalizes to Skills.}
Evaluating on \skillexb{} yields consistent gains across all models over the
original \skillret{}; for example, SkillRet-0.6B improves from 53.2 to 64.6
in NDCG@10 ($+$11.4\,pp), and zero-shot Qwen3-Embedding-0.6B rises from
49.9 to 60.8 ($+$10.9\,pp). This confirms that the one-to-one annotation
problem is not limited to tool retrieval.

\begin{figure}[t]
 \centering
 \includegraphics[width=\columnwidth]{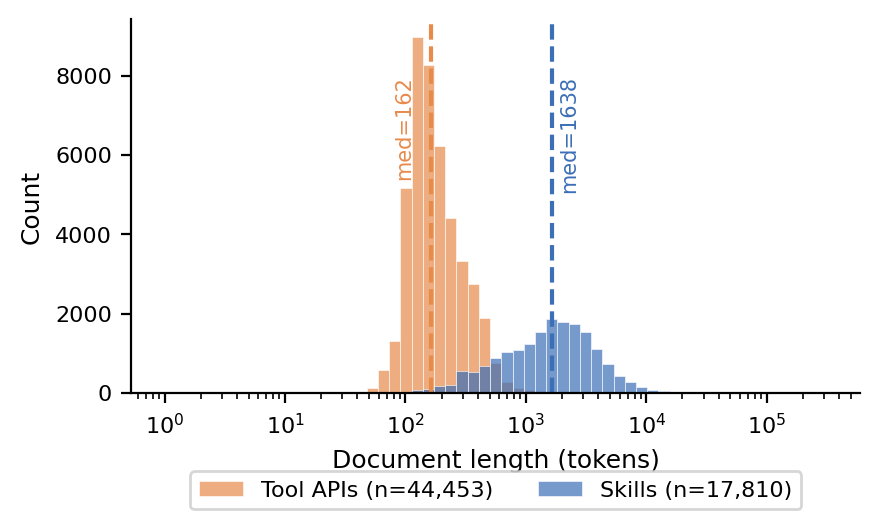}
 \caption{%
 Documentation length distribution for Tool API docs (orange) and Skill docs
 (blue).
 }
 \label{fig:doc_complexity}
\end{figure}

\paragraph{Skill Fine-Tuning Provides Genuine Improvement.}
Unlike tool retrieval, the fine-tuned SkillRet-8B still substantially
outperforms zero-shot Qwen3-8B \emph{after} expansion: 82.93 vs.\ 61.89
NDCG@10 ($+$21.0\,pp).
This contrast stands in sharp relief against the tool-retrieval result, where
the analogous gap collapses to $+$3.0\,pp. We hypothesize that the difference stems from documentation length (Figure~\ref{fig:doc_complexity}).
Skill documents contain rich, structured descriptions of capabilities,
invocation patterns, and examples; tool API documents are shorter and more
uniform.

\subsection{Downstream Task Evaluation on ToolBench}
\label{subsec:downstream}

\begin{figure}[t]
  \centering
  \includegraphics[width=\columnwidth]{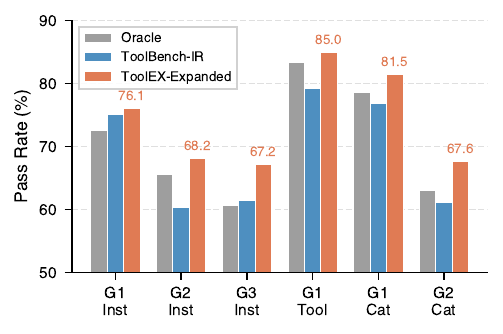}
  \caption{%
    Downstream task pass rates (\%) on ToolBench across six instruction categories. 
  }
  \label{fig:downstream}
\end{figure}

To test practical utility, we evaluate downstream task pass rates on 568
ToolBench tasks in three settings: \textbf{Oracle}, \textbf{ToolBench-IR},
and \textbf{ToolEX-expanded}, where the provided APIs are replaced by
functionally equivalent combinations discovered by our pipeline.
Figure~\ref{fig:downstream} reports results across six instruction categories.

ToolEX-expanded achieves the highest pass rate in all six categories,
exceeding Oracle by $+$3.5--6.5\,pp and ToolBench-IR by $+$1.0--6.9\,pp.
We view this as evidence that many discovered equivalents are practically
usable, not as proof that ToolEX surpasses a fully specified execution oracle.

\subsection{Discussion: Benchmark Implications}
\label{subsec:discussion}

\toolexb{} does not make evaluation more permissive; it restores credit for functionally
valid solutions already present in the repository but omitted by one-to-one
annotation. The gap between \toolde{} and \toolexb{} should therefore be
read as annotation-induced underestimation rather than as an arbitrary metric
shift. Exact-reference evaluation remains useful for reproducibility, but in
open-world tool ecosystems with substantial functional redundancy it should
not be treated as the only notion of correctness.

This distinction also affects how model gains are interpreted. Under a
single-reference benchmark, a retriever can be rewarded for matching the
annotated API identity while receiving no credit for an equally valid but
unannotated alternative. \toolexb{} instead evaluates \emph{functional
sufficiency}, which better matches deployment. The consistent gains on both
\skillexb{} and ToolBench suggest that the recovered credit reflects a broader
mismatch between one-to-one annotation and open-world tool use. We therefore
recommend reporting both exact-reference and equivalent-aware scores in future
tool-retrieval benchmarks.


\section{Conclusion}
\label{sec:conclusion}

We identify a one-to-many annotation problem in tool retrieval benchmarks:
functionally equivalent tools are common, but existing benchmarks annotate
only one ground-truth combination per query. \toolex{} addresses this with
an automated pipeline that expands equivalent positives. Under the resulting
\toolexb{} benchmark, 30--47\% of the apparent fine-tuning gain on
\toolde{} vanishes once equivalents are credited, showing that a substantial
part of the reported gain is an evaluation artifact. Extending the same
framework to SkillRet confirms that the problem is ecosystem-wide.

\paragraph{Limitations.}
(i) \toolex{} verifies tool equivalence semantically rather than by
practically invoking the tools, which may introduce false positives.
(ii) We conduct partial human spot-checking, but do not yet perform
exhaustive validation of the expanded labels at full scale.
more optimistic than conservative multi-reference alternatives.
(iii) While we have applied \toolex{} to \toolde{}, \skillret{}, and \toolret{},
generalization to other ecosystems remains to be validated.

\paragraph{Future Work.}
Extending \toolex{} to tool-use benchmarks where tools can be practically
executed would enable runtime verification of functional equivalence.
Active-learning loops could scale the pipeline to larger libraries by
prioritizing high-uncertainty candidates.
Finally, designing retrievers that return diverse equivalent sets opens a
new direction for tool-augmented agents~\citep{shen2024hugginggpt,patil2024gorilla}.

\bibliography{aaai2027}

\appendix
\section{LLM Prompt Templates}
\label{app:prompts}

\paragraph{Stage 1: Multi-Tool Query Decomposition.}
We use DeepSeek-V3.2 for query decomposition during benchmark construction.
The system prompt instructs the LLM to decompose a query into exactly $k$
atomic sub-queries, one per ground-truth tool, minimizing the semantic gap
between sub-query text and tool documentation (Listing~\ref{lst:stage1}).
Because this prompt receives the benchmark-provided tool count and reference
tools, it is an oracle-assisted annotation component rather than a deployable
test-time planner; the corresponding diagnostic rows in the main paper reuse it only
as an oracle upper bound.

\begin{prompttextbox}{Stage 1 --- Sub-query Decomposition (DeepSeek-V3.2)}
\refstepcounter{promptlisting}\label{lst:stage1}%
You are the Planner in a Plan-and-Execute system.

Given a user query and a set of relevant tools, your job is to decompose the query
into exactly N executable steps (subqueries), one per relevant tool, so that executing
all subqueries in order fully answers the query.

Each subquery will later be used as a search query to retrieve its corresponding tool
from a large tool library. Therefore, the primary goal of each subquery description is
to MINIMIZE THE SEMANTIC GAP between the subquery text and the tool documentation,
so that the correct tool can be accurately retrieved.

\medskip\textbf{Rules:}
1. Produce exactly N subqueries where N equals the number of relevant tools provided.
2. Each subquery must map to exactly one relevant tool via its "relevant\_tool\_id".
3. Write subquery descriptions with enough detail to enable accurate retrieval.
 Focus on two things only:
 (a) FUNCTIONALITY --- the precise operation the tool performs
 (e.g., retrieve, calculate, search, filter, authenticate, convert)
 (b) ENTITY --- the specific data subject or domain the operation acts on
 (e.g., stock ticker, historical events on a date, user credentials, ...)
 Use domain-relevant terminology that likely appears in the tool documentation.
 Do NOT describe specific parameter names, parameter types, or return value types.
4. Do NOT mention any tool name, API name, or function name. Describe the operation
 and data, not the implementation.

Return ONLY a valid JSON array --- no markdown fences, no explanation outside the JSON.
\end{prompttextbox}

\paragraph{Stage 2: Tool Verification.}
We use GPT-4o-mini to verify tool functionality for the sub-query. The system prompt instructs the LLM to judge whether a candidate tool can be
directly invoked to complete a given atomic sub-query, using the ground-truth
tool as a reference benchmark (Listing~\ref{lst:stage2}).

\begin{prompttextbox}{Stage 2 --- Tool Verification (GPT-4o-mini)}
\refstepcounter{promptlisting}\label{lst:stage2}%
Your task is to judge whether a given tool can be directly invoked to complete a given sub-query.

The sub-query is atomic --- it describes a single operation completable by calling one tool.
You are also given a Reference Tool --- a ground-truth tool that can already complete
this sub-query. If the tool under evaluation performs the same core operation or has equivalent
functionality, answer "yes".

\medskip\textbf{Rules:}
 Answer "yes" when ALL of the following hold:
 1. The tool's primary function directly performs the same operation as the Reference Tool.
 2. The tool's output covers what the sub-query needs (same type of result as the Reference Tool).

 Answer "no" when ANY of the following apply:
 1. The tool's function does not match the sub-query's required operation.
 2. The tool's output only partially satisfies the sub-query and significant extra steps
 are needed.

When genuinely uncertain, answer "no".

Return ONLY a valid JSON object:
\{"verdict": "yes" or "no", "reason": "<one concise sentence>"\}
\end{prompttextbox}

\paragraph{Stage 3: Combination Audit.}
We use Claude-Sonnet-5 to audit tool combinations for the specific query and instruction. The system prompt instructs the LLM to determine whether a set of available
tools is sufficient to fully complete the user's query, using a sub-query
breakdown as a coverage checklist and the ground-truth tool set as reference. In this stage, the LLM audits the provided tool combinations under the query context. 
(Listing~\ref{lst:stage3}).

\begin{prompttextbox}{Stage 3 --- Combination Audit}
\refstepcounter{promptlisting}\label{lst:stage3}%
You are a tool-use agent. Your task is to determine whether a given set of
available tools is sufficient to fully complete the user's query.

You will receive the user's query, an instruction providing sub-query context, and
the available tools. You may also receive a subquery breakdown --- use it as a
checklist to verify each required step is covered. If a reference tool set is
provided, treat it as a known-correct benchmark: tools that perform the same
core operations as the reference tools count as sufficient, even if they differ
in name, interface details, or parameter formats.

\medskip\textbf{Rules:}
1. Answer "yes" when the available tools collectively cover every operation needed
to fulfill the query and can produce the result the user wants. Note that
intermediate values (such as authentication tokens, resource IDs, or session
handles) are assumed to be passed between tools at runtime --- you do not need
to verify that explicitly; focus on whether the required operations exist.

2. Answer "no" when an essential operation is entirely absent from the available
tools --- meaning no tool can perform a required action at all (a functional
mismatch, not a minor format or naming difference).

Return ONLY a valid JSON object:
\{"verdict": "yes" or "no", "reason": "<one concise sentence>"\}
\end{prompttextbox}

\section{Human Validation Guidelines}
\label{app:human_validation}

This appendix specifies the protocol for the compact human validation used
to check automatically added positives from \toolex{}.

\paragraph{Sampling unit.}
We audit only the \emph{final equivalent tool-combination annotations},
i.e., accepted query-level combinations after Stage~3. In our implementation,
these items are stored in \texttt{query\_combo\_verified.jsonl}. Audit items
should be sampled stratified by domain (Code/Web/Customized) and by query
complexity (e.g., number of required tools / number of sub-queries).

\paragraph{Reviewer inputs.}
Each reviewer receives the original query, the task instruction, the
sub-query breakdown (if available), the candidate tool combination, the
reference ground-truth combination, and the model-produced rationale stored
with that combination.

\paragraph{Decision rule.}
Mark a candidate combination as \textbf{valid} iff both conditions hold:
(1)~the combination jointly covers all required operations in the original
query, and (2)~the tools are dependency-compatible whenever one tool's output
must feed another (e.g., shared platform, compatible identifiers, or session
state). Otherwise mark it \textbf{invalid}.

\paragraph{Adjudication and outputs.}
Each item is judged independently by two reviewers. Disagreements are
resolved by a third reviewer who sees both rationales but not model scores.
The audit should report combination-level precision, reviewer agreement, and
a small failure taxonomy with at least the following buckets: decomposition
ambiguity, near-synonym but non-substitutable tools, incomplete functional
coverage, and cross-platform dependency mismatch.

\section{Justification of Max-Aggregation}
\label{app:metric}

This section formalizes why $\max$-aggregation is the principled way
to extend a standard IR metric from one to many ground-truth
combinations, and why the gap $\Delta_M$ admits a clean interpretation.

\paragraph{Setup.}
Fix a query $q$, the retrieval list $L_q$, a cutoff $K$, and the expanded
set $\hat{\calC}_q=\{C_1,\dots,C_m\}$ of valid equivalent combinations.
Let $M(L_q,C_j)\in[0,1]$ denote any of $\ndcg@K$, $\recall@K$, $\comp@K$
computed against the single reference $C_j$ under the same single-reference
definitions used in the main paper.
All combinations in $\hat{\calC}_q$ are by construction \emph{equally
valid} solutions to $q$: the pipeline verifies each to be functionally
sufficient, and none is privileged over another.
The one-to-one benchmark resolves this equivalence arbitrarily by selecting
one $C^*\in\hat{\calC}_q$ as the sole reference; \toolexb{} instead asks the
model-independent question: \emph{among all valid solutions, how well does
the retrieval list do?}

\paragraph{Principled multi-reference extension.}
Each $M(L_q,C_j)$ is the score the retrieval earns if $C_j$ were the
reference. Because every $C_j$ is an equally valid reference, the only
label-agnostic score---one that does not depend on which combination the
annotator happened to record---is a symmetric aggregate over
$\{M(L_q,C_j)\}_{j=1}^{m}$.
The $\max$ selects the most favorable valid reference and is the unique
choice consistent with the semantics ``a retrieval is correct if it matches
\emph{any} valid equivalent combination.''
It is equivalently the indicator of the event ``the retrieval list contains
some combination that, taken as a reference, achieves the top score,''
reducing to the standard single-reference metric when $|\hat{\calC}_q|=1$.
Aggregates such as the mean would instead discount retrievals that match
some but not all equivalents, re-introducing a false-negative penalty; the
$\max$ alone credits each valid match in full.

\paragraph{Upper bound and gap.}
Since Stage~3 guarantees $C^*\in\hat{\calC}_q$, the $\max$ ranges over a
superset of the one-to-one reference, giving
$M^*(q)=\max_j M(L_q,C_j)\ge M(L_q,C^*)$ for every metric.
Thus \toolexb{} upper-bounds \toolde{} query-by-query and metric-by-metric,
and the gap $\Delta_M(q)=M^*(q)-M(L_q,C^*)\ge 0$ is exactly the score
\toolexb{} restores by recognizing a retrieved equivalent that the
one-to-one benchmark counted as a miss.
When the retrieval already places $C^*$ at least as well as any other
equivalent, $\Delta_M(q)=0$; $\Delta_M(q)>0$ precisely when some
unannotated equivalent $C_j\neq C^*$ yields a higher single-reference score
than $C^*$, i.e., when the one-to-one convention caused an underestimate.
Because $\Delta_M\ge 0$ holds pointwise, it also holds after
macro-averaging, so the aggregate gap is a non-negative measure of
annotation-induced underestimation, separable from genuine retrieval ability.

\paragraph{Per-metric independence.}
The $\max$ is applied independently to each metric
to each metric independently, so the reference combination
realizing $\max_j\ndcg@K$, $\max_j\recall@K$, and $\max_j\comp@K$ may
differ for the same query. This is deliberate: the three metrics reward different aspects
of the ranking (rank-weighted quality, top-$K$ coverage, full coverage), and
forcing a single proxy combination to optimize all three would distort at
least one. 


\end{document}